\documentstyle[11pt]{article}
\def\psnormal{\textwidth=16cm\textheight=22cm
\oddsidemargin=0.5cm\evensidemargin=0cm
\parindent=1cm}

\psnormal
\newcommand{\be}{\begin{equation}}
\newcommand{\ee}{\end{equation}}

\begin{document}

\begin{titlepage}
\begin{flushright}
OUTP-9209P \\
CERN-TH.6562/92\\
July 1992
\end{flushright}
\vspace{2.0cm}
\begin{centering}

{\LARGE\bf Phase Structure of QED$_3$ at Finite Temperature}

\vspace{1.50cm}
{I.J.R. Aitchison$^{(a)}$, N. Dorey$^{(b)}$,
M. Klein-Kreisler$^{(a)}$ and N.E. Mavromatos$^{(c)}$}

\vspace{1.00cm}
{\small
$^{(a)}$ Department of
Theoretical Physics, 1 Keble Rd, Oxford OX1 3NP, UK.\\
$^{(b)}$ Theoretical Division T-8, MS
B285, Los Alamos National Laboratory,\\
Los Alamos, NM 87545, USA.\\
$^{(c)}$ Theory Division, CERN, CH-1211, Geneva 23, Switzerland.\\ }
\date{1992}

\def\theequation{\arabic{equation}}
\vspace{3.00cm}
\baselineskip=16pt

\begin{abstract}
{\def\baselinestretch{1.0}
\noindent Dynamical symmetry breaking in three-dimensional QED with N
fermion flavours
is considered at finite temperature, in the large $N$ approximation.
Using an approximate treatment of the Schwinger-Dyson equation for the
fermion self-energy, we find that chiral symmetry is restored above a
certain critical temperature
which depends itself on $N$. We find that the ratio of the zero-momentum
zero-temperature fermion mass to the critical temperature has a large
value compared with four-fermion theories, as had been suggested in a
previous work with a momentum-independent self-energy.
Evidence of a temperature-dependent critical $N$ is shown to appear in
this approximation. The phase diagram for spontaneous mass generation
in the theory is presented in $T-N$ space.}
\end{abstract}

\end{centering}

\end{titlepage}
\newpage

\def\baselinestretch{1.5}
\pagenumbering{arabic}

\section{\bf Introduction}
\vspace{1.00cm}

\indent  Quantum Electrodynamics in (2+1) dimensions
(QED$_3$) has attracted considerable interest in the
last few years \cite{Pis} - \cite{Atkin}.
One reason for this is that it provides a
simple setting for the study of dynamical chiral symmetry breaking
which is important for theories such as QCD. Furthermore QED$_3$
also appears to be relevant to some long-wavelength models
of certain 2D condensed matter systems,
including high-T$_c$ superconductors \cite{Dor,Dor3}.
At zero temperature a number of studies have shown that chiral symmetry
is dynamically broken in  QED$_3$.
Using the leading order in the $1/N$ expansion of the Schwinger-Dyson (SD)
equations
Appelquist et al. \cite{App} showed that the theory exhibited a critical
behaviour as the number $N$ of fermion flavours approached
$N_c=32/\pi^2$; that is, a fermion mass was dynamically generated only
for $N<N_c$. Qualitatively the same
behaviour was found by Nash \cite{Nash},
who included $O(1/N^2)$ corrections. As against this, Pennington and
collaborators \cite{Penn}, adopting
a more general non-perturbative approach
to the SD equations, found that the dynamically generated fermion mass
decreased exponentially with $N$, vanishing only as $N\rightarrow\infty$
(as originally found by Pisarski \cite{Pis} using a simplified form of
of the SD equation). On the other hand, an alternative non-perturbative
study by Atkinson et al. \cite{Atkin} suggested that chiral symmetry is
indeed unbroken at sufficiently large $N$. We should also mention that
Pisarski \cite{Pis2} has used the renormalization group approach and
the $\epsilon$ expansion to argue that chiral symmetry remains broken
for all $N$, but this result has to be interpreted with some caution
as the relevant value of $\epsilon$ is 1.
 The theory has also been simulated
 on the lattice \cite{Dag} and the results
appear to be consistent with the existence of a critical $N$ as predicted
in the analysis of ref \cite{App}. On the other hand, because Monte-Carlo
simulations are performed on  lattices of finite size $L$ and typically
cannot detect mass scales less than the IR cutoff scale $1/L$, the
persistence of an exponentially small fermion mass for large $N$, as
suggested in \cite{Penn}, cannot be ruled out.

The extension of the above type of analysis to finite temperature is
extremely important, and highly relevant to either application already
mentioned. The most obvious question is whether there is a critical
temperature $T_c$ above which chiral symmetry is restored.
This was first answered affirmatively by Kocic \cite{Kocic}, using a
very simple approximation to the finite temperature SD equations,
in which the entropy of the fermions was not fully taken into account.
An improved calculation was made by Dorey and Mavromatos \cite{Dor2},
based on the finite temperature Schwinger-Dyson equations, to leading
order in $1/N$. These authors found that the ratio $r$ of twice the
zero-temperature mass to the critical temperature was approximately
independent of $N$ and much larger $(r\simeq 10)$ than the value obtained
in (BCS-like) four fermion theories $(r$ at most $3.5)$.
The latter result
could be relevant to an understanding of this ratio in the high-$T_c$
superconductors. In ref \cite{Dor2}, however, the
(leading order in $1/N$)
SD equation for the fermion self-energy function
$\Sigma(p)$ was considerably
simplified by making the assumption \cite{Pis}
that $\Sigma$ was in fact a
constant, independent of $p$. It is clearly necessary to be assured that
the results of \cite{Dor2} are essentially
independent of this assumption.

In the present paper we extend and complete
the analysis of Ref. \cite{Dor2}.
We again start from the Schwinger-Dyson equation
for $\Sigma(p)$, to leading
order in $1/N$, but keep the full momentum dependence of $\Sigma$.
Following the approach of \cite{Dor2}, and of similar
studies of chiral symmetry
restoration in
four dimensions \cite{Davis}, we further simplify the SD equation
by
adopting
an instantaneous approximation for the kernel. This corresponds to
retaining
only the part of the kernel corresponding to the static interaction
between charges. We
find that the main conclusions of Ref. \cite{Dor2} are
indeed robust --in particular,
the ratio $r$ has almost the same value.

A second question concerns the
existence of a critical number of flavours $N_c$,
above which chiral symmetry would be restored. Within
the $1/N$ approach, there
is no $N_c$ if $\Sigma$ is approximated
by a constant \cite{Pis}, as already
noted, a result which naturally persists at finite $T$
in the approximation
\cite{Dor2} ; but with a $p$-dependent $\Sigma$ we find a (temperature
dependent) $N_c$, analogous to Ref. \cite{App}. We are
thus able to present
the resulting phase structure of the model, as a function
of the variables
$N$ and $T$: there is a single critical line separating the region of low
$(N,T)$ where $\Sigma\neq 0$  from the region of high $(N,T)$ where
$\Sigma=0$ (see Figure 7 below).
As indicated above, we are aware
that criticism have been raised \cite{Penn}
concerning the reliability of the
$1/N$ approach for calculating fermion masses
$\Sigma$ well below the intrinsic scale set by the dimensionful coupling
$\alpha$ (as happens in this case) and it may be that in the exact theory
no sharply defined $N_c$ exists (though it would seem very difficult to
establish
this conclusively, given that {\it some} approximation to the complete
set of SD equations has to be made). However, at the very least,
the lattice
results at $T=0$ strongly indicate the existence of an effective
critical {\it region}, in which the dynamical fermion mass undergoes a
rapid crossover from a regime of large values to one of very small values
as $N$ is increased. We shall take the view here that our
$1/N$-based phase diagram should give a qualitatively correct picture
in this sense --and that in any case there is merit in obtaining as
complete
an analysis as possible of one definite model at finite temperature.
It would clearly be of interest to investigate the effect, at finite
temperature, of the non-perturbative vertex structure advocated in Ref.
\cite{Penn}.

\vspace{1.00cm}

\section{\bf Momentum-dependent self-energy (gap) equation
          at finite temperature}
\vspace{0.5cm}

The Lagrangian for massless QED$_3$ with $N$ flavours is
\be
{\cal L}=-{1\over 4}f_{\mu\nu}f^{\mu\nu}+\overline{\psi}_{i}
(i\!\not\!\partial -e\!\not\! a)\psi_{i}
\label{Lag}
\ee
where $a_{\mu}$ is the vector potential, $i=1,2,...N$,
and a reducible four dimensional Dirac
algebra has been chosen so that (\ref{Lag}) has a
continuous chiral symmetry. The conventions of
\cite{App} will be adopted throughout.

The Schwinger-Dyson equation for the fermion propagator at non-zero
temperature $k_B T=\beta^{-1}$ is given by
\be
S_{F}^{-1}(p_{0},P,\beta) = S_{F}^{(0)-1} (p)-
               {{e}\over {\beta}}
               \sum_{n=-\infty}^{+\infty}
               \int{{d^2{\bf k}}\over {(2\pi)^2}}
               \gamma^{\mu} S_{F} (k_{0},K,\beta)
               \Delta_{\mu\nu} (q_{0},Q,\beta)
               \Gamma_{\beta}^{\mu}
\label{SD}
\ee
where
\be
\begin{array}{lcl}
p=(p_{0},\bf{p}) & P=|\bf{p}| & p_{0}=(2m+1){\pi/
\beta} \\
k=(k_{0},\bf{k}) & K=|\bf{k}| & k_{0}=(2n+1){\pi/
\beta} \\
q=(q_{0},\bf{q}) & Q=|\bf{q}|=|\bf{p-k}| & q_{0}=2(m-
n){\pi/\beta}
\end{array}
\label{momenta}
\ee
As stated above,
we truncate (\ref{SD}) by working at leading order in $1/N$,
in which case
$\Gamma^{\nu}$ is replaced by its bare value $e\gamma^{\nu}$ and
$\Delta_{\mu\nu}$ by  the $O(1/N)$ propagator shown in
Figure 1, in
which the fermions are massless (the massless vacuum polarisation
loop already softens the photon propagator \cite{Pis}, and the exact form
of this softening does not qualitatively change the behaviour of the
fermion propagator \cite{Penn}). We shall work in Landau gauge and assume
that the wave
function renormalisation can be neglected to leading order in
$1/N$, so that
$S_{F}^{-1}={\not\!{p}} + \Sigma_{m}(P,\beta)$ (note however the criticism
made of this
step by Pennington and Walsh \cite{Penn}).
The trace of
equation (\ref{SD}) then yields a closed integral equation for
$\Sigma_{m}$:
\be
\Sigma_{m} (p)= {\alpha\over N\beta} \sum_{n=-
\infty}^{\infty} \int {{d^2{\bf k}}\over {(2\pi)^{2}}}
          \Delta (q_{0},Q,\beta)
          {{\Sigma_{n} (K,\beta)}\over
          {k^{2}+\Sigma_{n}^{2}(K,\beta)}},
\label{SDbeta}
\ee
\noindent with
\be
\Delta(q_{0},Q,\beta)={1\over 8}
Tr[\gamma^{\mu}\Delta_{\mu\nu} (q_{0},Q,\beta)\gamma^{\nu}]
\label{ladelta}
\ee
and $\alpha=Ne^2$. We now follow Ref. \cite{Dor2} in retaining only the
$\mu=\nu=0$ component of $\Delta_{\mu\nu}$ at zero frequency and thus set,

\be
 \Delta_{\mu\nu}(q_{0},Q,\beta)=
{{\delta_{\mu 0}\delta_{0\nu}}\over {Q^{2} +\Pi_{0}(Q,\beta)}}
\label{Delta0}
\ee
\noindent where \cite{Dor2}
\be
\Pi_{0} (Q,\beta)={{2\alpha}\over
                    {\pi\beta}}\int^{1}_{0}
                    dx \,\ln(2\,\cosh ({{Q\beta}\over {2}}
                    \sqrt{x(1-x)})).
\label{Pi0}
\ee

In this approximation
$\Sigma_{m}(P)$ becomes
frequency  independent and the summation over $n$ in
(\ref{SDbeta}) can be performed analytically yielding
\be
\Sigma (P,\beta)= {\alpha\over {8N\pi^{2}}} \int d^2{\bf k}
{{\Sigma (K,\beta)}\over {Q^{2} +\Pi_{0} (Q,\beta)}}
{{\tanh\,{\beta\over 2} \sqrt{K^{2} +\Sigma^{2}(K,\beta)}}
\over \sqrt{K^{2} +\Sigma^{2}(K,\beta)}}.
\label{GapEqn}
\ee
\noindent The main purpose of this paper is to present the detailed
analysis of this {\it temperature- and momentum-dependent} gap equation.

\vspace{1.00cm}

\section{\bf Numerical procedure and results}

Equation (\ref{GapEqn}) involves
a two-dimensional integral over ${\bf k}$,
while $\Pi_0$ in (\ref{Pi0})
involves $Q\!=\!|{\bf p}\!-\!{\bf k}|$ inside a
further integral. The analogous $T=0$ equation has frequently been
simplified \cite{Kog} by
replacing $Q$ by $max(K,P)$, rendering the angular
integral trivial. We do not
make this approximation here, but we have found it
very convenient to adopt an excellent analytic approximation to (\ref{Pi0})
(correct to about 1.5\% at worst) which is provided by the expression
\be
\Pi_0 (Q,\beta) ={\alpha\over {8\beta}} \left[
                    Q\beta+ {16\ln2\over\pi}\,
               \exp\,(-{\pi\over{16\ln2}} Q\beta)\right],
\label{Approx}
\ee
\noindent which incorporates the correct limiting behaviour as
either $Q$ or $\beta$ tends to zero or infinity. In particular, as
$Q\rightarrow 0$,
\be
\Pi_0 (Q,\beta) \rightarrow {{2\alpha\ln 2}\over {\pi\beta}}
\label{Pi0Q0}
\ee
\noindent which exhibits the thermal screening noted before \cite{Dor2}.

As regards to the integral over $K=|{\bf k}|$ in (\ref{GapEqn}), this
of course extends to $K\rightarrow\infty$ in principle. For numerical
purposes the upper limit would normally be replaced by some cutoff
parameter $\Lambda$, chosen to be sufficiently large that further
increase of it makes no difference. In the present case however,
the dimensionful parameter $\alpha$ provides a natural scale: in
particular, Appelquist et al. \cite{App} noted that the integral in the
corresponding $T=0$ equation was rapidly damped for momenta greater than
$\alpha$, so
that effectively $\Lambda\simeq\alpha$. We have found that the
same is true for equation (\ref{GapEqn}) at finite temperature, and thus
from now on we shall work with (\ref{GapEqn}) cut off at
$\Lambda=\alpha$, and present our results in terms of the scaled quantity
$(\Sigma/\alpha)$, momenta being also scaled by $\alpha$.

We have used a numerical algorithm to solve equation
(\ref{GapEqn}). An iterative procedure has been followed
to find a solution for $\Sigma(P,\beta)$. As usual the
algorithm only converged if the input function was sufficiently
close to the true solution. The iteration was started by
adopting the approximation $Q\simeq max(K,P)$ and solving
the resulting version of (\ref{GapEqn}) for a large value of
$\beta$. This solution then provided the input trial function
for the true equation (\ref{GapEqn}) at the same low temperature.
When the input and the output of the integral equation
agreed within a 2\% of difference the iterative sequence was
stopped, and the output was considered as the solution
to equation (\ref{GapEqn}) for the defined temperature.
$\beta$ was then incremented downwards in small steps using
the solution of (\ref{GapEqn}) for the previous $\beta$ as the
input function for the next. In this way we were able to
obtain the dependence of $\Sigma$ on $\beta$, as well as on $p$.

The scaled dynamical mass as a function of scaled
momentum is shown in Figure 2 for $N=1$, at various temperatures.
We notice that the mass
remains constant for a wide range of momenta up to
roughly $P\simeq \Sigma(P=0)$ and then rapidly drops to zero.
We also notice that as expected the mass decreases with
rising temperature. The rate
at which this occurs grows as we approach the critical
temperature $T_c$, above which the mass vanishes (and which will
depend on $N$). This is illustrated in Figure 3, which shows the zero
momentum mass $\Sigma(0,\beta)/\alpha$ as a function of the scaled
temperature $k_BT/\alpha$
for $N=1,1.5$. The approach of $\Sigma(0,\beta)$
to zero in the vicinity of $T_c$ can be studied by plotting
$\ln\Sigma$ versus $\ln (T_c-T)$; we find that $\Sigma\simeq (T_c-T)^x$
where the exponent $x$ depends on $N$ and lies between  0.4 and 0.6 for
$N$ between 1 and 2. This is consistent (up to subleading $N$-dependent
corrections) with the value $x=1/2$ characteristic of BCS theory.

Figure 4 shows the scaled mass versus scaled momentum for several
values of $N$, at a fixed value of $\beta$. It is clear that $\Sigma$
decreases strongly as $N$ increases from $N=1$, suggesting that it may
vanish for $N$ larger than some critical $N_c$. This possibility
is examined in Figure 5, which shows $\Sigma(0,\beta)/\alpha$
versus $N$ for various values of $\beta$. We are not able to follow
the $\Sigma/\alpha$ curves much below values of order $10^{-5}$ due
to numerical difficulties, but it seems reasonable to
conclude that at a given temperature $T$, $\Sigma$ indeed vanishes
for $N>N_c$, where $N_c$ depends on $T$. For large $\beta$
(low $T$), $N_c$ approaches a value just greater than 2. As the
temperature is raised, $N_c$ decreases. At $T\!=\!0$ Appelquist
et al. \cite{App} have found that in the limit $N\rightarrow N_c$,
the zero-momentum mass vanishes according to
\be
{\Sigma\over\alpha} \propto{\exp\left[-2\pi\over{\sqrt{N_c/N-1}}\right]}.
\label{NcAp}
\ee
We have explored the possibility of a similar behaviour in our model.
As convergence near the critical point is very slow we have had to
extrapolate from the calculated values of $N$ to get the critical
value $N_c$ for the corresponding temperature. Our results are shown
in Figure 6 where we have plotted $-\ln (\Sigma (0,\beta)/\alpha)$
vs. $1/\sqrt{N_c(T)/N-1}$; we observe that for fixed temperature the
curves approach  straight lines as $N$ approaches $N_c$. This leads us to
believe that indeed in this region the zero-momentum mass behaves like
\be
{{\Sigma (0,T)} \over\alpha}
 \propto{\exp\left[-C(T)\over{\sqrt{N_c(T)/N-1}}\right]}.
\label{NcUs}
\ee
for some temperature-dependent function $C(T)$ (which is, however, at the
temperatures shown in Figure 6, considerably smaller in magnitude than
the $T=0$ value of $2\pi$ given in (\ref{NcAp})).

As mentioned
in the previous paragraph, we find that as $T\rightarrow 0$
$N_c(T)$ approaches
a value just greater than 2. This is, of course, different
from the $N_c$ found
in Ref. \cite{App},
using an equation which ought to be the
zero temperature limit
of ours, and hence some further comment is required.
In fact, while it is
obviously true that the $T\rightarrow 0$ limit of the
{\it full} SD
equation (\ref{SD}) must be the same as that of Ref. \cite{App},
this is not the case after the instantaneous approximation has been made,
leading to Eqns.(\ref{SDbeta}) and (\ref{GapEqn}). Nevertheless, Eqn.
(\ref{SDbeta}) does reduce as $T\rightarrow 0$ to an equation of similar
form (for small $\Sigma$) to that in Ref. \cite{App}, but the numerical
coefficient in front of the integral is a factor 1.5--2.0 times too small.
Effectively this means that in comparing our results with those of
Ref. \cite{App} we should take our $N$ as being roughly equivalent to
the $N$ of Ref. \cite{App} divided by this factor. This is the reason
for the discrepancy in the $N_c(T=0)$ values.

These results enable us to obtain the phase diagram shown in Figure 7.
There is a single critical line, such that for $(N,T)$ below this line
$\Sigma\neq 0$, and for $(N,T)$ above it $\Sigma=0$. We have only shown
 the region $N\geq 1$, but it seems likely that the line approaches
$N=0$ asymptotically as $T\rightarrow\infty$. In this plot we have
rescaled the critical line to match the zero-temperature results of
Ref. \cite{App}, namely $N_c(T=0)=3.2$.

As mentioned in the Introduction, the dimensionless ratio
$r=2\Sigma(P\!=\!0,T\!=\!0)/k_B T_c$ is an important quantity,
distinguishing between different mass-generation mechanisms.
Our values of $r$ are shown in Table 1 for $N=1,1.5,1.7$, where
for convenience we also list the corresponding values of
$\Sigma(P\!=\!0,T\!=\!0)/\alpha$ and of $k_B T_c /\alpha$.
This table can be directly compared with Table 1 of Ref. \cite{Dor2}
which --it will be recalled-- was obtained by solving a simplified
equation in which the $P$-dependence of $\Sigma$ was neglected.
The comparison shows that while our more exact equation (\ref{GapEqn})
yields values of the mass $\Sigma$ which are about one order of
magnitude smaller than those obtained in Ref. \cite{Dor2}, the critical
temperature $T_c$ is also correspondingly reduced, so that $r$
remains with a value of order 10, in agreement with the value
found in \cite{Dor2}, and also approximately independent of N.

With an eye to the possible relevance to high-$T_c$ superconductivity,
it is natural to wonder about the orders of magnitudes of the quantities
appearing in our results, when expressed in physical units. We must
emphasize, however, that relatively small changes in the kernel (which
is after all only an approximation) can make rather large changes in
$\Sigma$. It is
clear that the scale of the model is set by the dimensionful
quantity $\alpha$, which has dimensions of $(length)^{-1}$ or $(energy)$
in the system in
which we have implicitly been working, namely that in which
$\hbar\!=\!v\!=\!1$ with $v$ the Fermi velocity in the condensed matter
case (and that of light for QED$_3$). If $\alpha$ is regarded as a freely
adjustable parameter, then from Table 1 estimating $k_BT_c\sim 10^{-4}
\alpha$ for $N=2$
(the value required by our high-$T_c$ model \cite{Dor3}),
we would
need to take $\alpha\sim 80 \:eV$ in order to obtain $T_c\sim 100^oK$
as is
required experimentally. It seems hard to understand how such a large
energy
could arise naturally. If we readjusted the ``effective'' $N-T$ curve
so as
to agree with $N_c(T=0)$ as found in Ref. \cite{App}, we would obtain
$k_BT_c\sim 10^{-3} \alpha$ for $N=2$, leading to a required $\alpha$
of order $8\:eV$. These different estimates merely underline,
of course, the
difficulty in making anything other than rather rough order
of magnitude calculations as far as numerical values are concerned.
A more reliable determination of $\Sigma$ and $T_{c}$ would be obtained
from finite temperature Monte-Carlo simulation of lattice QED$_{3}$.

A value of $\alpha$ in the region of a few $eV$ is still much larger than
typical Heisenberg
exchange energies (recall that in our model \cite{Dor3}
the gauge field arises
in connection with the spin degrees of freedom in the
original lattice Hamiltonian). Nevertheless,
it is possible to form
an estimate of $e^2$ in terms of the parameters of the
lattice model of \cite{Dor3}, which
shows that it is effectively enhanced,
as follows. The lattice analogue of the fermion
kinetic energy is the ``hopping term'' which enters with coefficient $t$.
If the lattice fermion operators are rescaled by $ta$ (where $a$ is the
lattice spacing) so as to get the correct dimensions of the fields in the
continuum limit, and if space is then
rescaled so as to obtain the (Dirac)
kinetic energy with unit coefficient, the lattice $U(1)$ coupling $g$
becomes effectively replaced by $g/(ta)^{1/2}$. An
estimate of $ta$ in such
models may be obtained by noting that according to
Baskaran et al. \cite{Bask}
the maximum doping concentration $n_{max}\sim t/U$, where $U$ is the
Hubbard repulsion. Since we may take $U\sim a^{-1}$ ($i.e.\; U\rightarrow
\infty$ in the continuum limit) we find $ta\sim n_{max}$, which has the
empirical value of only a few percent. Assuming that the magnitude of
$g$ is set by the spin magnitude (1/2), and its
length scale by the lattice
spacing $a$, we obtain finally for the square of
the effective coupling
\be
e^2 \sim {1\over {4a}} {{\hbar v}\over{n_{max}}}
\label{e2}
\ee
having
reinstated $\hbar$ and $v$. The latter
quantity can be conveniently
found  from the relation $\xi\sim\hbar v/\Sigma$ for the correlation
length $\xi$. Using $\xi\sim 30$\AA and $\Sigma\sim 5k_BT_c$ we find
$v/c\sim 5\times 10^{-4}$, which gives
$e^2 \sim$ few $eV$. Thus it is perhaps
not impossible that such values could arise within the context of a model
such as that of Ref. \cite{Dor3}.

\vspace{3.00cm}
\noindent{\large\bf Acknowledgements}

We thank Mike Pennington for a very useful discussion about the content
of the works listed in Ref. \cite{Penn}.
MKK wishes to thank the National University of Mexico, the SERC and
Los Alamos National Laboratory for financial support.

\newpage

\begin{figure}
\noindent{\Large\bf Tables}
\begin{center}
\begin{tabular}{|c||c|c|c|}   \hline
 {$N$} & {$1$} & {$1.5$} & {$1.7$} \\ \hline
 $\Sigma(P=0,T=0)/\Lambda$ & $1.93\times 10^{-2}$ & $2.6\times 10^{-3}$
& $9.0\times 10^{-4}$              \\ \hline
 $k_{B}T_{c}/\Lambda$  & $3.76 \times 10^{-3}$  & $5.2\times 10^{-4}$
& $1.90 \times 10^{-4}$     \\ \hline
 $r=2\Sigma(P=0,T=0)/k_{B}T_{c}$ & $10.26$ & $10.0$ & $9.5$ \\ \hline
\end{tabular} \end{center}
{\def\baselinestretch{1}
Table 1: The zero-temperature and zero-momentum fermion mass and the
critical temperature at $\alpha/\Lambda=1$ with the ratio $r$ for
$N=1,1.5,1.7$.}
\end{figure}
\vspace*{2.50cm}

\noindent{\Large\bf Figure captions}
\vspace{1.00cm}

\noindent Figure 1: The photon propagator to leading order in $1/N$

\noindent Figure 2: Scaled dynamical mass $\Sigma/\alpha$ as a function
of scaled momentum $p/\alpha$ for $N=1$, and various scaled (inverse)
temperatures $\beta\alpha$.

\noindent Figure 3: Zero-momentum scaled mass versus scaled temperature
for $N=1,1.5$.

\noindent Figure 4: Scaled mass versus scaled momentum for $\beta\alpha=
10^5$ for $N=1,1.2,1.5$.

\noindent Figure 5: Zero-momentum scaled mass versus $N$ at various (inverse)
temperatures.

\noindent Figure 6: Test for the behaviour given by Eqn. (\ref{NcUs}) near
the critical region.

\noindent Figure 7: Phase diagram for spontaneous mass generation.

\end{document}